\documentclass[useAMS,usenatbib]{mn2e}
\usepackage{epsfig}
\usepackage{epstopdf}
\usepackage{graphicx}
\usepackage{amsmath}
\usepackage{amssymb}
\usepackage{natbib}
\newcommand{\msun}{\ensuremath{{\rm M_{\odot}}}}

\newcommand{\apj}{ApJ}
\newcommand{\apjl}{ApJ}
\newcommand{\mnras}{MNRAS}
\newcommand{\aj}{AJ}
\newcommand{\apjs}{ApJS}
\newcommand{\nat}{Nat}

\newcommand{\aap}{A\&A}

\newcommand{\physrep}{Physics Reports}

\newcommand{\jcap}{JCAP}

\begin{document}

\title[Faint AGN in $z\gtrsim6$ LBGs]{Faint AGN in $z \gtrsim 6$ Lyman-break Galaxies Powered by Cold Accretion and Rapid Angular Momentum Transport}

\author[Mu{\~n}oz and Furlanetto]
{
Joseph A.\ Mu{\~n}oz\thanks{E-mail:jamunoz@astro.ucla.edu} 
and 
Steven Furlanetto
\\University of California Los Angeles, Department of Physics and Astronomy; Los Angeles, CA 90095, USA
}

\maketitle

\begin{abstract}
We develop a radiation pressure-balanced model for the interstellar medium of high-redshift galaxies that describes many facets of galaxy formation at $z \gtrsim 6$, including star formation rates and distributions and gas accretion onto central black holes.  We first show that the vertical gravitational force in the disk of such a model is dominated by the disk self-gravity supported by the radiation pressure of ionizing starlight on gas.  Constraining our model to reproduce the UV luminosity function of Lyman-break galaxies (LBGs), we limit the available parameter-space to wind mass-loading factors 1--4 times the canonical value for momentum-driven winds.  We then focus our study by exploring the effects of different angular momentum transport mechanisms in the galactic disk and find that accretion driven by gravitational torques, such as from linear spiral waves or non-linear orbit crossings, can build up black hole masses by $z=6$ consistent with the canonical $M$-$\sigma$ relation with a duty cycle of unity, while accretion mediated by a local viscosity such as in an $\alpha$-disk results in negligible BH accretion.  Both gravitational torque models produce X-ray emission from active galactic nuclei (AGN) in high-redshift LBGs in excess of the estimated contribution from high-mass X-ray binaries.  Using a recent analysis of deep {\it {Chandra}} observations by \citeauthor{Cowie12}, we can already begin to rule out the most extreme regions of our parameter-space: the inflow velocity of gas through the disk must either be less than one percent of the disk circular velocity or the X-ray luminosity of the AGN must be substantially obscured.  Moderately deeper future observations or larger sample sizes will be able to probe the more reasonable range of angular momentum transport models and obscuring geometries.

\end{abstract}


%-------------------------------------------------------------------------------------------------------------
%             Introduction
%-------------------------------------------------------------------------------------------------------------
\section{Introduction}

Luminous quasars have recently been observed out to $z \gtrsim 6$ \citep[e.g.,][]{Fan06, Mortlock11} indicating the presence of rapidly accreting supermassive black holes (BHs) in the early universe.  These active galactic nuclei (AGN) are thought to be hosted by extremely rare and massive dark matter halos with masses around $10^{12}\,\msun$ undergoing frequent major mergers that drive gas toward the galactic center \citep{Li07}.  

At the same time, Lyman-break galaxies (LBGs) have been discovered at $z \gtrsim 6$ \citep[e.g.,][]{Bouwens06, Bunker10, McLure10, Finkelstein10, Bouwens11a, Bouwens11b} with typical star formation rates (SFRs) of $\lesssim 1\,{\rm \msun/yr}$.  An analysis of the resulting UV luminosity functions (LFs) shows that these LBGs inhabit $\sim10^{10}\,\msun$ halos \citep{Munoz12} that are much more common than those that host high-redshift AGN.  It is still uncertain whether they produce enough high-energy photons either to reionize the universe or to keep it ionized \citep[e.g.,][]{ML11, Bouwens12}.  Yet, for the most part, these galaxies are too faint for efficient high-resolution spectra and too small to be fully resolved \citep{Oesch10b, WL11}.  Thus, observations have, so far, revealed little about their interstellar media (ISM).

In the absence of such data, it is often assumed that $z \gtrsim 6$ LBGs are physically similar to the lower-redshift starbursts that have also been found with the Lyman-break technique.  The fact that few $z\sim2$ LBGs host AGN \citep[e.g.,][]{Steidel04} is often pointed to as evidence that few $z\sim6$ galaxies should contain supermassive BHs as well \citep[e.g.,][]{Cowie12}.  Indeed, studies of X-ray heating in the intergalactic medium (IGM) \citep[e.g.,][]{Oh01, Furlanetto06} have cited the low-redshift AGN LF to argue that AGN are not necessary for cosmic reionization and that high-mass X-ray binaries (HMXBs) provide the only significant sources of X-rays.  These studies are limited by data that may be missing faint AGN in $z\gtrsim6$ galaxies.

On the other hand, a general theoretical picture of galaxy physics is emerging where galactic accretion is balanced by star formation, winds, and angular momentum transport in marginally Toomre-unstable disks \citep[e.g.,][]{Thompson05, CAFG11, HQ11, Hopkins12a, Dave12} and in which gas inflow toward the center of the galactic center is a generic prediction \citep{HQ10, HQ11, Bournaud11}.  Moreover, recent analysis of {\it{Chandra}} observations has begun to place limits on the X-ray luminosity of $z \gtrsim 6$ LBGs using a stacking method \citep{Treister11, Cowie12}.  While the results of different studies have been inconsistent, the conclusions by \citet{Treister11} may reflect evidence of ongoing and highly obscured black hole growth in faint AGN at the centers of these high-redshift LBGs.  However, predictions for such X-ray observations based on the new paradigms of galaxy formation are absent in the literature.  With a view toward resolving this shortfall, we start from the radiation pressure-balanced ISM model of \citet[][hereafter TQM05]{Thompson05} that was shown to describe massive star-forming galaxies and AGN at $z\sim2$ and extend it to even higher redshift by assuming a dust-free ISM and applying the appropriate velocity dispersions and metallicities.  We improve the treatment by considering both the self-gravity of the disk and the ejection of material through winds and show how the UV mass-to-light ratio and the X-ray luminosity are signatures of and place constraints on the galaxy fueling rate, wind properties, and angular momentum transport mechanism.  

In \S\ref{sec:model:acc_disk:model}, we outline the components of our ISM model showing explicitly how it differs from previous studies and implementations and clearly laying out the free parameters.  After constraining the wind mass-loading parameter to reproduce observations of the UV LF at $z \gtrsim 6$ in \S\ref{sec:UVLF}, we explore the effects of different angular momentum transport models on the central black hole accretion rate in \S\ref{sec:BHacc} and on the resulting AGN luminosity and its obscuration by the dense gas in \S\ref{sec:agn}.  We present our results in \S\ref{sec:results} where we specifically consider the X-ray emission of high-redshift LBGs, comparing our model AGN outputs with expectations from HMXBs and the recent {\it{Chandra}} observations.  Finally, we conclude with a review and discussion of our conclusions in \S\ref{sec:conclusions}.

%-------------------------------------------------------------------------------------------------------------
%             Model
%-------------------------------------------------------------------------------------------------------------
\section{The Model}\label{sec:model:acc_disk:model}
%-------------------------------------------------------------------------------------------------------------
\subsection{Overview}\label{sec:model:acc_disk:model:overview}
In this subsection, we summarize the basic components of the model discussed in the rest of \S\ref{sec:model:acc_disk:model} with fiducial values and free parameters.  We formulate this model with the goal of representing $z \gtrsim 6$ LBGs in mind.  At lower redshift, it reduces roughly to the model that TQM05 showed accurately describes luminous AGN and starbursts at $z\sim2$.  Here we also point out where our formulation differs from that in previous studies.

We start with the basic model of TQM05 with a galactic disk that connects smoothly to an AGN accretion disk around a central BH.  We note that, while the disks need not be thin, this picture assumes that high-redsift galaxies are not spherically amorphous.  The size and velocity dispersion of the galactic disk are specified by the mass and redshift of the host dark matter halo.  As described in \S\ref{sec:model:acc_disk:model:disk}, gas is accreted onto the outer edge and, at each radius, may either be turned into stars, ejected by winds, or transported toward smaller radii.  Given their focus on larger galaxies, TQM05 ignored the loss of gas via winds.  However, we include a momentum-driven outflow at each radius with a mass loss rate proportional to the ratio of SFR to velocity dispersion and normalized by the free parameter $\eta_{0}$.  The timescale for the buildup of the disk is long enough that the reservoir of existing gas need not be included either as contributing to star formation or as a sink for accreting gas.  Ultimately, the fraction of gas not incorporated into stars or ejected by winds is accreted onto the BH.  

The rate of baryonic accretion onto the galaxy (\S\ref{sec:model:acc_disk:model:acc_global}) and the mechanism for gas transport through the disk (\S\ref{sec:model:acc_disk:model:acc_disk}) are key elements of the model.  We assume that the cold-flow model---in which infalling cold gas is not shocked at the virial radius but streams freely to the galaxy---gives the most reasonable description of the galactic fueling rate in $z \gtrsim 6$ LBGs and is completely specified by halo mass and redshift.  Moreover, we consider three possible mechanisms for the transport of gas through the disk: (i) an $\alpha$-disk model with a viscosity parameterized by a constant value $\alpha$; (ii) a gravitational torque-generated infall model induced by a linear spiral wave and characterized by an infall velocity that reaches a constant fraction, $m$, of the local sound speed; and (iii) a shocked infall model induced by orbit crossings of spiral waves in which the infall velocity is a fixed fraction, $\beta$, of the orbital velocity.  This last process was not considered by TQM05 but has been seen in numerical simulations and developed analytically \citep{HQ10, HQ11}.  In each case, we take fiducial values of $\alpha=0.3$, $m=0.2$, or $\beta=0.01$.

Given the global accretion rate, an angular momentum transport mechanism, and a wind parameter $\eta_{0}$, we can calculate the density and temperature profiles as well as the star formation and infall rates as a function of radius in the disk by solving the equation of vertical hydrostatic equilibrium\footnote{Following TQM05, where multiple solutions exist, we select the one at the lowest temperature and ignore the multiplicity of phases in the ISM that these other solutions might imply.} (\S\ref{sec:model:pbal}), in which radiation pressure from stars dominates thermal pressure from supernovae and maintains marginal Toomre-stability.  While we consider the effect of additional turbulent support, either from dense clumps in cold streams \citep{Dekel09b} or disk instabilities \citep{Burkert10, KB10, Bournaud11}, we ultimately ignore both when compared with pressure from stars (\S\ref{sec:model:turb}).  Deviating from the TQM05 model, we include the self-gravity of the disk in the pressure balance and assume a dust-free ISM, taking as negligible additional support from pressure on dust grains (\S\ref{sec:model:dust}).  

%-------------------------------------------------------------------------------------------------------------
\subsection{Balancing Inflows, Outflows, and Star Formation}\label{sec:model:acc_disk:model:disk}
The TQM05 model assumes a galactic disk that extends toward small radii and matches onto the accretion disk around a central BH.  This system is embedded in an isothermal halo with velocity dispersion $\sigma$ so that the angular rotation speed of the disk as a function of radius is given by
\begin{equation}\label{eq:OMK}
\Omega = \sqrt{\frac{G\,M_{\rm BH}}{r^3}+\frac{2\,\sigma^2}{r^2}},
\end{equation}
where $M_{\rm BH}$ is the mass of the BH.  Given the paucity of high-redshift data and the uncertainty in extrapolating the local $M$-$\sigma$ relation to early times, we simply assume the local empirical relation 
\begin{equation}\label{eq:Mbh}
M_{\rm BH}=2\times10^{8}\,\sigma_{200}^4\,\msun
\end{equation}
with $\sigma_{200}=\sigma/(200\,{\rm km/s})$ \citep{Tremaine02} and substitute the total halo velocity dispersion for the bulge dispersion.  For a $10^{10}\,\msun$ halo at $z=6$, where $\sigma \approx 50\,{\rm km/s}$, the extra contribution to the rotational velocity from the BH is significant only within the central parsec; throughout most of the disk, $\Omega \approx \sqrt{2}\,\sigma/r$.  

We assume that gas in the galactic disk is optically thick to the ionizing UV radiation emitted by stars so that radiation pressure maintains vertical hydrostatic equilibrium against gravity.  In the outer regions where pressure from stars dominates, the disk is constrained to be marginally Toomre-stable \citep{Toomre64} with 
\begin{equation}\label{eq:Q}
Q=\frac{\sqrt{2}\,\Omega\,c_{\rm s}}{\pi\,G\,\Sigma_{\rm g}}=1
\end{equation}
where $c_{\rm s}$ is the total sound speed in the disk, $\Sigma_{\rm g}=2\,\rho\,h$ is the gas surface density, $\rho$ is the gas density in the plane, and the disk scale height is $h$.  Thermal gas pressure and radiation from the AGN are allowed to stabilize the inner disk to $Q>1$.  We review this pressure equilibrium in \S\ref{sec:model:pbal}, but we refer the reader to TQM05 for more details.  

The model implicitly assumes that the timescale for building up the reservoir of gas in the disk is much longer than any of the disk crossing timescale, the star formation timescale, or the advection timescale.  Thus, none of the gas accreting onto the disk goes toward increasing the gas mass of the disk.  For momentum-driven winds, this is true for galaxies hosted in halos less massive than $10^{11}\,\msun$ out to $z=10$ \citep{Dave12}.  Instead, accreted gas can be turned into stars or expelled by winds, as considered by \citet{Dave12}, but a small fraction may also be accreted onto the central BH and power an AGN.  This conservation is also true on a local level as gas is transported through the disk, so that the inflow rate at each radius is:
\begin{equation}\label{eq:Mdot1}
\dot{M}(r)=\dot{M}_{\rm disk}-\int_r^{R_{\rm disk}}\,2\,\pi\,r'\,\dot{\Sigma}_{\star}\,(1+\eta_{\rm wind})\,dr',
\end{equation}
where $\dot{M}_{\rm disk}$ is the galactic accretion rate of baryons onto the outer edge of the disk at a radius of $R_{\rm disk}$, $\dot{\Sigma}_{\star}$ is the surface density of star formation, and $\eta_{\rm wind} \equiv \dot{M}_{\rm wind}/\dot{M}_{\star}$ is the wind mass-loading parameter.  We take $R_{\rm disk}$ to be a fixed fraction $\lambda/\sqrt{2}$ of the halo virial radius, $R_{\rm vir}$, with $\lambda=0.05$ \citep{MMW98}, consistent with the highest-resolution observations available \citep{Oesch10b, WL11}.  While TQM05 ignored the reduction in gas accretion rate through outflows, we include the effect of a momentum-driven wind \citep{Murray05} with
\begin{equation}\label{eq:eta_wind}
\eta_{\rm wind}=\eta_{0}\,\frac{100\,{\rm km/s}}{\sigma}\,\frac{\epsilon}{10^{-3}},
\end{equation}
where $\eta_{0}$ is a free parameter to which we will return in \S\ref{sec:UVLF} and $\epsilon$ is the efficiency with which stars process matter into radiation.  In principle, $\epsilon$ depends on the IMF and metallicity of the stars in the galaxy with an efficiency of $\epsilon_0 \sim 10^{-3}$ for a Salpeter IMF from 1--100$\,\msun$ at solar metallicity.  However, using Starburst99 \citep{Leitherer99}, we have checked that $\epsilon$ differs from $\epsilon_0$ by only a factor of order unity even for a Pop III IMF at very low metallicities.  Thus, we will assume $\epsilon=\epsilon_0$ throughout the rest of this {\it{Paper}}.\footnote{\label{foot:epsilon1}Since only ionizing radiation from stars impacts the dust-free ISM in our model, the value of $\epsilon$ may be correspondingly reduced.  However, because of the degeneracy between $\epsilon$ and $\eta_{0}$ in equation \ref{eq:eta_wind}, this ultimately has no effect on $\eta_{\rm wind}$ itself (see footnote \ref{foot:epsilon2}).}  Given this efficiency, if the asymptotic wind velocity at infinity is $\approx 3\,\sigma$ \citep{Murray05}, the critical velocity below which winds reduce the gas accretion rate more than star formation itself is $300\,\eta_{0}\,{\rm km/s}$, implying that most high-redshift galaxies are strongly affected by winds.

On the other hand, the semi-analytic models of \citet{Raicevic10} and \citet{Lacey11} postulate that LBGs at $z \gtrsim 6$ are the products of mergers, which are probably not well-represented by the steady-state model we have assumed in equations \ref{eq:OMK} and \ref{eq:Mdot1}.  At $z=6$, the halo merger time of $10^{10}\,\msun$ halos is on the order of the Hubble time \citep[e.g.,][]{KL12}, but the small sizes and high densities of early galaxies may suppress merger-generated effects \citep{Cen12}.  Given the complexities of such a model, we leave a detailed analysis of this scenario to future work.   
 
The galactic accretion rate of baryons onto the disk, $\dot{M}_{\rm disk}$, is a key component of equation \ref{eq:Mdot1} since it limits the amount of possible star formation, winds, and AGN activity.  By contrast, \citet{Krumholz09b} proposes a star formation efficiency set by the properties of molecular clouds.  This picture can be reconciled with our model if the molecular gas fraction is simply set by the SFR required to maintain $Q\sim1$ \citep{Hopkins11}.  In \S\ref{sec:model:acc_disk:model:acc_global}, we consider the galactic accretion rate due to cold flows.  The transport of the gas through the disk is also a critical component of the model since the transport rate determines whether gas will form stars and expel winds (i.e., if the rate is slow) or be accreted onto a luminous AGN (i.e., if the gas accretes quickly through the disk so that the advection timescale is shorter than the star formation timescale).  \S\ref{sec:model:acc_disk:model:acc_disk} considers three different mechanisms for the transport of angular momentum and gas through the disk and their implications.

%-------------------------------------------------------------------------------------------------------------
\subsection{Galactic Gas Accretion}\label{sec:model:acc_disk:model:acc_global}
The accretion rate of baryons onto the disk is an important parameter in our model because it sets the amount of gas available for star formation and AGN fueling.  As we will see in \S\ref{sec:model:acc_disk:model:acc_disk}, the accretion rate also determines the scale height required to channel such a mass flow rate through the disk at a certain velocity.  

In the cold flow model, baryons stream directly onto the galactic disk and are not slowed by an accretion shock \citep{Keres05, Dekel09a}.  Thus, the accretion rate roughly traces the buildup of dark matter in the halo.  Several studies have investigated fitting formulae for the average accretion rate as a function of only halo mass, $M_{\rm halo}$, and redshift \citep[e.g.,][]{Neistein06, McBride09, CAFG11}.  While all of these studies were focused on results at $z\sim2$, we assume that the same redshift dependence holds out to $z\gtrsim6$ and gives a reasonable approximation of the baryonic accretion rate at these redshifts.  Following the work by \citet{McBride09}, the average baryon accretion rate in cold flows at high redshift is:
\begin{equation}\label{eq:acc_cf} 
\dot{M}_{\rm disk} \approx 3\,{\rm \msun/yr}\,\left(\frac{M_{\rm halo}}{10^{10}\,\msun}\right)^{1.127}\,\left(\frac{1+z}{7}\right)^{2.5}\,\left(\frac{f_{\rm b}}{0.16}\right),
\end{equation}
where $f_{\rm b}$ is the cosmic baryon fraction.  We note that this prescription would be significantly altered in a merger scenario, where gas may be dumped onto galaxies at much higher rates for short periods of time and at chaotic intervals \citep[e.g.,][]{Li07}.

%-------------------------------------------------------------------------------------------------------------
\subsection{Angular Momentum Transport Through the Disk}\label{sec:model:acc_disk:model:acc_disk}

The rate at which gas is transported toward the center of the disk determines how much is consumed by star formation and how much fuels the AGN.  We consider three different mechanisms that may operate in the disks of high-redshift galaxies: a simple $\alpha$-disk model, infall mediated by linear spiral waves, and non-linear shocked infall due to orbit crossings.  In each case, the mass infall rate as a function of radius is given by
\begin{equation}\label{eq:Mdot2}
\dot{M}=4\,\pi\,r\,h\,\rho\,v_{\rm in},
\end{equation}
where $v_{\rm in}$ is the infall velocity of the gas as a function of radius.  With $\rho$ determined by equation \ref{eq:Q}, the mechanism that sets $v_{\rm in}$ also indirectly determines the disk scale height required maintain mass conservation while channeling a given $\dot{M}$ through the disk.

\subsubsection{$\alpha$-disk}\label{sec:model:acc_disk:model:acc_disk:alpha}
In an $\alpha$-disk, the viscosity funneling gas into the center of the disk is a local process given by $\nu=\alpha\,c_{\rm s}\,h$ and described by the parameter $\alpha$ \citep{SS73}.  Following TWM05, the infall velocity is
\begin{equation}\label{eq:vin_a}
v_{\rm in}=\nu\,\left|\frac{d{\rm ln}\Omega}{dr}\right|.
\end{equation}
Unless otherwise noted, we take a fiducial value of $\alpha=0.3$.  However, the inflow can achieve even lower values of $\alpha$ in certain regimes \citep{Gammie01}.  Since a smaller $\alpha$ would only reduce the already miniscule BH accretion rates we find for locally viscous disks (see \S\ref{sec:BHacc}), we consider only the higher value here.  For a thin disk, the infall velocity of an $\alpha$-disk is much less than the sound speed.

\subsubsection{Linear Spiral Wave}\label{sec:model:acc_disk:m}
Since a local mechanism for viscosity may have trouble fueling the central BHs of luminous AGN, TQM05 further consider a global-torque model in which the infall velocity reaches a fixed fraction of the sound speed \citep[e.g][]{Goodman03}\footnote{It may be the case that BHs are fed by the small fraction of gas with initially low angular momentum \citep{Goodman03} rather than by gas whose angular momentum has been transported outward.  We leave an exploration of this possibility and its incorporation into our models for future work.}:
\begin{equation}\label{eq:vin_m}
v_{\rm in}=m\,c_{\rm s},
\end{equation}
where $m$ is the Mach number.  This type of angular momentum transport is produced by gravitational torques due to linear spiral waves (LSWs) in pure gas disks \citep[e.g.,][see also \citealt{HQ11} and references therein]{Kalnajs71, LBK72}.  Following \citet{Goodman03} and TQM05, we assume a fiducial value of $m=0.2$.  Thus, the infall velocity in this model is a significant fraction of the sound speed, even for a thin disk.  As we will see, this allows more gas to accrete onto the BH than in the $\alpha$-disk case.

\subsubsection{Stellar Torques and Shocked Infall}\label{sec:model:acc_disk:shock}
Two additional effects will lead to increased angular momentum transport and gas infall in galactic disks.  First, in disks with both gas and stars, wave modes in each component will be offset by $\sim10\,{\rm deg}$ due to dissipation in the collisional gas \citep{Noguchi88, BH96, Berentzen07}.  The resulting gravitational torque of the stars on the gas drives gas toward the center.  However this effect is strongly suppressed in systems where the disk does not dominate the potential \citep{HQ10, HQ11}.  Additionally, nonlinear wave modes in the gas lead to orbit crossings \citep[e.g.,][]{PP77}, orbit trapping \citep[e.g.,]{BT87}, and ultimately shocked dissipation and infall.  Both of these processes lead to mass infall rates that are independent of the local sound speed and, thus, of the local thermodynamics of the disk \citep{HQ11}.  In general,
\begin{equation}\label{eq:vin_s}
v_{\rm in}=\beta\,r\,\Omega,
\end{equation}
where $\beta$ is a constant.  Because we assumed in \S\ref{sec:model:acc_disk:model:disk} that the disk is a subdominant component of the potential in $z \gtrsim 6$ galaxies, we do not expect a stellar disk to provide a significant torque on the gas.  Therefore, we will refer to the angular momentum transport mechanism of equation \ref{eq:vin_s} as resulting from ``shocked infall."  On the other hand, the formation of disks at the highest redshifts is an open problem.  Numerical simulations with significant inflow that include both effects find typical values of $\beta \sim 0.001$--$0.1$ \citep{HQ10}.  In what follows, we nominally assume $\beta=0.01$ and consider results for a range of values where appropriate.

%-------------------------------------------------------------------------------------------------------------
\subsection{Pressure Balance}\label{sec:model:pbal}

We assume that hydrostatic equilibrium is maintained vertically in the disk.  For a thin disk, the pressure required to balance gravity can be approximated as:
\begin{equation}\label{eq:pgrav}
P_{\rm grav}=2\,\pi\,G\,\Sigma_{\rm g}\,\rho\,h+\rho\,h^2\,\Omega^2,
\end{equation}
where the first term on the right-hand-side is the contribution from the self-gravity of the disk and the second is due to the isothermal halo potential.  We show that, for constant $Q$, equation \ref{eq:pgrav} can also be expressed as $P_{\rm grav}=B\,\rho\,h^2\,\Omega^2$, where $B$ is a constant.  While TQM05 assumes $B=1$ for simplicity, essentially ignoring the disk self-gravity term, we show that the contribution of this term is indeed significant.  Since $c_{\rm s}^2=P_{\rm grav}/\rho$, then $c_{\rm s}=B^{1/2}\,h\,\Omega$.  Substituting into equation \ref{eq:Q} with $\Sigma_{\rm g}=2\,\rho\,h$ and solving for the density gives
\begin{equation}\label{eq:rho}
\rho=\frac{B^{\frac{1}{2}}\,\Omega^2}{\sqrt{2}\,\pi\,G\,Q},
\end{equation}
which is different from the TQM05 derivation by the factor of $B^{1/2}$.  Substituting back into equation \ref{eq:pgrav} yields $P_{\rm grav}=(2^{3/2}\,B^{1/2}/Q+1)\,\rho\,h^2\,\Omega^2=B\,\rho\,h^2\,\Omega^2$ from which we find (for $Q=1$) $B=(\sqrt{2}+\sqrt{3})^2 \approx 9.9$.  While such a large value of $B$ implies that the disk self-gravity is an important component of $P_{\rm grav}$, this is only because the gravitational pull of the halo contributes only its small vertical component.  The potential and total gravitational force throughout most of the disk is still dominated by the isothermal halo as in equation \ref{eq:OMK}.

Gravity in the disk is balanced by radiation pressure from stars and the thermal pressure of the gas.  TQM05 considered the influence of radiation on grains within a coupled dust-gas medium.  However, in the dust-free ISM that we assume for $z\gtrsim6$ LBGs, UV photons impart momentum directly to the gas through ionizations.  In this case, the gas surface densities required for optical thickness are extremely small ($\sim10^{-5}\,{\rm \msun/yr}$).  For an ionizing escape fraction $f_{\rm esc}\ll1$, the resulting pressure is of the same order as that in the single-scattering case of a dusty medium, where the dust is optically thick to UV radiation but optically thin to the re-radiated IR \citep[e.g.,][]{AT11}, and given by:
\begin{equation}\label{eq:pstars}
P_{\rm stars, rad}=\epsilon\,\dot{\Sigma}_{\star}\,c.
\end{equation}
Note that we have, additionally, assumed no cancelation of oppositely-directed momentum \citep[e.g.,][]{Socrates08} in equation~\ref{eq:pstars}. 

TQM05 showed that the contribution to the pressure from a hot ISM phase in supernova bubbles is negligible because of the high density of luminous starbursts.  However, the mechanical deposition of energy from gas swept into shells by supernova shocks provides pressure support of the same order as $P_{\rm stars, rad}$.  The same argument holds true in the comparably high densities of the early galaxies we consider here.  Thus, we set the total pressure from stars to be $P_{\rm stars}= 2\,P_{\rm stars, rad}$ and have checked that this assumption provides nearly identical results as explicitly including mechanical pressure support from supernovae.  Because contraction of the disk leads to an increase in star formation while expansion leads to a decrease, pressure due to star formation is a self-regulating process\footnote{Here we implicitly assume the ``continuum" approximation discussed in footnotes 12 and 14 of TQM05 so that star formation feedback continues to maintain $Q=1$ even in annuli where the SFR is very small and discreteness effects become important.} that maintains $Q \sim 1$ in the disk.

While the disk is dense enough on average to gain momentum from every ionizing photon produced, winds or the AGN may create holes in the gas distribution through which radiation can leak and increase $f_{\rm esc}$.  In this case, stabilizing the disk will require additional star formation and result in less gas available for accretion into the central AGN.  However, we would expect the resulting leakage to have only a small effect on the mechanical pressure supplied by supernovae.  Therefore, holes should not change $P_{\rm stars}$ by more than a factor of two.  The same argument can be applied to using more precise values of $\epsilon$ for pressure generated exclusively by ionizing radiation.

Thermal gas pressure,
\begin{equation}\label{eq:pgas}
P_{\rm gas}=\frac{\rho\,k_{\rm B}\,T}{m_{\rm p}},
\end{equation}
also provides support against gravity according to its density and temperature.  Here, $m_{\rm p}$ is the proton mass, and, in the star-forming region of the disk
\begin{equation}
\sigma_{\rm SB}\,T^4 = \frac{1}{2}\,\epsilon\,\dot{\Sigma}_{\star}\,c^2
\end{equation}
so that the mid-plane temperature, $T$, is approximately the effective temperature of the radiation.\footnote{We note that our dust-free formulation differs from the treatment in TQM05 and \citet{SG03} where the mid-plane temperature $T^4=T^4_{\rm eff}\,(3\,\tau/4+\tau^{-1}/2+1)$, where $\tau$ is the IR optical depth.  Additionally, a lower value of $\epsilon$ would not significantly affect the gas temperature because the corresponding reduction in pressure would be raised to the $1/4$ power.}  We also ignore the effect of CMB heating on the gas.  The CMB temperature at $z=6$ is about $19\,{\rm K}$, which can be higher than our modeled gas temperature at some radii and in some regions of parameter space.  However, the Compton heating time for the CMB is long, and formally, there are no metal lines through which it can interact with the gas in our dust-free model.  Still, the CMB may have an important effect on the temperature of high-redshift molecular clouds and certainly on our ability to observe radio lines from these objects \citep{Obreschkow09b}.  

Vertical hydrostatic equilibrium implies 
\begin{equation}\label{eq:pbal}
P_{\rm grav} = P_{\rm stars} + P_{\rm gas}.
\end{equation}
As in the TQM05 treatment, once $P_{\rm gas}$ comes to dominate, star formation is no longer required to support the disk and becomes negligible.  As gas accretes to even lower radii, it loses the gravitational energy that powers the AGN.  However, since star formation can no longer deplete the supply of gas to the central BH, we ignore the question of how and whether this extra energy heats or supports the disk.

%-------------------------------------------------------------------------------------------------------------
\subsection{Turbulence from Dense Gas Clumps and Disk Instabilities}\label{sec:model:turb}

In this subsection, we consider turbulent pressure support due to the impact of dense clumps in a cold flow on the disk \citep{Dekel09b} or instabilities within the disk \citep{Burkert10, KB10, Bournaud11}, which have been postulated to be the dominant mechanisms maintaining $Q=1$.

Since the angular momentum of cold flows is thought to be responsible for the buildup of rotating disks \citep{Dekel09b, Stewart11}, we assume that dense clumps within the streams impact and deposit their kinetic energy as turbulence into the disk at its edge.  This turbulence is transported inward over the advection timescale $\sim R_{\rm disk}/v_{\rm in}$ but decays over the disk crossing time $\sim h/c_{\rm s}$.  For an LSW model of gas transport (Eq. \ref{eq:vin_m}), the decay time is a fraction $\sim m\,h/R_{\rm disk}$ of the advection time and even smaller in an $\alpha$-disk model.  In a shocked infall model, this ratio is $\sim \beta\,\sigma\,h/(c_{\rm s}\,R_{\rm disk})$.  Thus, for a thin disk, $m \ll 1$, or $\beta \ll 1$, cold clump turbulence decays quickly and does not affect the disk beyond its outermost rim.

A related source of turbulent disk support may come from the energy released by the gas as it spirals inward through the disk and down the potential gradient.  If all of this energy is quickly radiated, then pressure of this magnitude is sub-dominant compared to stars, except very close to the galactic center.  Otherwise, if the turbulence remains un-thermalized, it may dominate and stabilize the disk as several studies that ignore pressure from stars claim in the literature \citep[e.g.][]{Burkert10, KB10, Bournaud11}.  Formally, this pressure may suppress all star formation in our model, but such an occurrence depends sensitively on, e.g., the details of the turbulent dissipation.  However, dominant instability-driven turbulence is in tension with other works that show star formation as participating directly in disk support and being regulated by the amount of cold inflow into the galaxy \citep[e.g.][]{Hopkins11, Dave12}.  Our goal is not to definitively settle this question, which must be left for numerical simulations, but since our work implicitly assumes the latter theoretical framework, we here ignore the contribution of lost gravitational energy to the turbulent support of the disk.

Finally, we note that additional turbulence in a major merger scenario--not considered here--may maintain $Q>1$ and allow gas to bypass star formation and funnel rapidly into a central, bright AGN \citep[e.g.,][]{Li07}.

%-------------------------------------------------------------------------------------------------------------
\subsection{Dust Content and Metallicity}\label{sec:model:dust}

Our treatment has assumed that the ISM of $z\gtrsim6$ galaxies are dust-free\footnote{There may, if fact, be significant amounts of dust confined to the cores of giant molecular clouds where star formation takes place and on which radiation pressure can act to support the clouds against gravitational collapse \citep{Murray10}.  However, this does not contradict our dust-free assumption, which applies to the large-scale regime where we have taken the gas and star formation to be smooth and homogeneous at a given radius.  A lack of dust-mixing may be a natural consequence of the density of high-redsift galaxies into which supernova bubbles have difficulty expanding (see \S\ref{sec:model:pbal}).}.  However, the dust content and metallicity of high redshift galaxies is not yet certain and widely debated.  While metals and dust have clearly been observed in $z \sim 6$ quasars \citep[e.g.,][]{Wang07, Wang11a, Wang11b}, the blue spectral slopes of LBGs indicate that they may be dust-free \citep[][but see \citet{Dunlop12}]{Bouwens10a}.  Theoretically, some studies postulate metal-free population III stars \citep[e.g.,][]{TS09, Trenti09, Johnson10, Cai11, Zackrisson11} or top-heavy IMFs \citep[e.g.,][]{Raicevic10, Lacey11, Kim11} in galaxies at $z \gtrsim 6$.  Others find that normal stellar populations are sufficient to reproduce the measured UV LF but disagree about the amount of dust extinction required \citep{Salvaterra11, Finlator11}.

For our purposes, whether momentum is supplied to the gas directly from ionizing radiation or via UV incident on dust grains, the resulting radiation pressure will not be significantly different from that given in equation \ref{eq:pstars} as long as the dust is optically thin to the re-radiated IR.  In luminous AGN and starbursts, this reprocessed radiation can provide additional pressure support for maintaining hydrostatic equilibrium \citep[][TQM05]{SG03}, increasing $P_{\rm stars, rad}$ by a factor of $(\tau+1)$, where $\tau=\kappa\,\Sigma_{\rm g}/2$ is the optical depth of dust to IR radiation and $\kappa$ is the Rosseland mean opacity in the optically thick limit.  Since $\kappa$ gives the absorption cross-section per unit {\emph{gas}} mass, it is ultimately $\kappa$ that reflects the assumed metallicity and dust-to-gas ratio of the system.  Despite the debate surrounding the exact dust content of $z \gtrsim 6$ galaxies, it is likely that they have significantly lower dust opacities than the ones appropriate for protoplanetary disks in the Milky Way \citep{BL94, Semenov03} and used in TQM05.

Our dust-free approximation is further justified by the small velocity dispersions and infall rates of interest.  For angular momentum transport mediated by LSWs, combining equations \ref{eq:Mdot2} and \ref{eq:vin_m} yields
\begin{equation}\label{eq:Sigma_g}
\Sigma_{\rm g}=\left(\frac{\sigma\,\dot{M}}{\pi^2\,G\,Q\,m\,r^2}\right)^{\frac{1}{2}}.
\end{equation}
Assuming the simple, low-temperature opacity model used by TQM05 where $\kappa=\kappa_{0}\,T_{\rm d}^2$, with dust temperature $T_{\rm d}$ and $\kappa_{0} \approx 2.4\times10^{-4}\,{\rm cm^2\,g^{-1}\,K^{-2}}$, the resulting optical depth is
\begin{equation}\label{eq:tau}
\tau \approx 4.2\,\left(\frac{\sigma}{300\,{\rm km/s}}\right)^{\frac{1}{2}}\,\left(\frac{\dot{M}}{320\,{\rm \msun/yr}}\right)^{\frac{1}{2}}\,\left(\frac{r}{200\,{\rm pc}}\right)^{-1}\,\left(\frac{T_{\rm d}}{100\,{\rm K}}\right)^{2}.
\end{equation}
While the luminous starbursts and AGN considered by TQM05 have $\sigma \sim 300\,{\rm km/s}$ and $\dot{M} \sim 300\,{\rm \msun/yr}$ at the edge of the disk, $z=6$ LBGs are significantly smaller systems with $\sigma \approx 50\,{\rm km/s}$ and a cold-flow accretion rate of $\dot{M} \approx 3\,{\rm \msun/yr}$.  This results in a much lower IR optical depth, even for similar opacities.  Repeating the argument for shocked infall and scaling to values appropriate for high-redshift LBGs, the optical depth is 
\begin{equation}\label{eq:tau_s}
\tau \approx 0.8\,\left(\frac{\sigma}{50\,{\rm km/s}}\right)^{-1}\,\left(\frac{\dot{M}}{3\,{\rm \msun/yr}}\right)\,\left(\frac{r}{200\,{\rm pc}}\right)^{-1}\,\left(\frac{T_{\rm d}}{100\,{\rm K}}\right)^{2}.
\end{equation}
While, in this case, $\tau$ is of order unity at the edge of the disk, we will see in \S\ref{sec:BHacc} that $\dot{M}/r$ drops rapidly just inside the edge, so that $\tau \ll 1$.

The uncertainty in the dust content of $z \gtrsim 6$ galaxies is also an issue for determining the relationship between the intrinsic UV radiation (at $1500\,${\rm \AA}) from star formation and that observed.  In \S\ref{sec:UVLF}, we consider two possibilities for the amount of UV dust extinction present in high-z galaxies as we constrain the wind parameter $\eta_{0}$.  However, we anticipate that upcoming observations with the Atacama Large Millimeter Array (ALMA) will shed more light on this point.

%-------------------------------------------------------------------------------------------------------------
%             Matching the LF
%-------------------------------------------------------------------------------------------------------------
\section{Constraints from the UV LF}\label{sec:UVLF}

The model in \S\ref{sec:model:acc_disk:model} sets the average star formation and AGN accretion rates for galaxies hosted in halos of a given mass and redshift.  In this section, we use current measurements of the UV LF to (1) determine the appropriate halo masses corresponding to $z=6$ LBGs and (2) constrain the wind parameter $\eta_{0}$.  This will ensure that our model galaxies will have the abundances and total star formation rates observed.

%-------------------------------------------------------------------------------------------------------------
\subsection{Average M/L}\label{sec:UVLF:M-L}

\citet{ML11} and \citet{Munoz12} inferred the evolution in the mass-to-light ratio, M/L, based on UV LF observations.  In their model, each galaxy has a distribution of possible luminosities with probabilities depending only on the mass of the host halo given by $dP(L_{1500}|M_{\rm halo})/d{\rm log}L_{1500}$.  Here, we invert this relation using the Bayesian method, to obtain the probability distribution of halo masses associated with a specific luminosity at $1500\,${\rm \AA}, $L_{1500}$:
\begin{equation}\label{eq:Mdist}
\frac{dP(M_{\rm halo}|L_{1500})}{d{\rm log}M_{\rm halo}} \propto \epsilon_{\rm AF}(M_{\rm halo}) \, \frac{dn(M_{\rm halo})}{dM_{\rm halo}} \,\frac{dP(L_{1500}|M_{\rm halo})}{d{\rm log}L_{1500}}.
\end{equation}
Here, $dn/dM_{\rm halo}$ is the halo mass function, and $\epsilon_{\rm AF}$ is the fraction of halos that host galaxies and depends on a critical suppression mass, $M_{\rm supp}$.  Since larger halos are built up from smaller ones, $\epsilon_{\rm AF}<1$ even for some halo masses larger than $M_{\rm supp}$.  However, if the minimum halo mass capable of hosting a galaxy is $M_{\rm supp}=10^{8}\,\msun$, then $\epsilon_{\rm AF} \approx 1$ for all currently observed $z\approx6$ LBGs.  Fitting the model to the UV LF of {\it{i}}-dropouts at $z\approx6$ \citep{Bouwens07} reveals that $dP(L_{1500}|M_{\rm halo})/d{\rm log}L_{1500}$ is a log-normal distribution with mean given by
\begin{equation}\label{eq:Lave}
\left<{\rm log}L_{1500}\right>\approx 27.19\,\left(\frac{M_{\rm halo}}{10^{10}\,\msun}\right)
\end{equation}
at $z=6$ and roughly constant standard deviation $\sigma_{L} \approx 0.25$ \citep{ML11, Munoz12}.  

Using this model, the average halo mass corresponding to the faintest HUDF {\it{i}}-dropout with an absolute magnitude of $M_{\rm AB}=-17.3$ \citep{Bouwens07} is approximately $10^{10}\,\msun$ with a distribution in log-space having a standard deviation of $\sigma_{\rm m} \approx 0.24$.  At $z=6$, this average mass corresponds to a velocity dispersion of $\sigma \approx 46\,{\rm km/s}$.  The mass of the halo hosting each galaxy specifies the average baryon accretion rate for a given accretion model.  It also sets the black hole mass based on our assumed $M$-$\sigma$ relation (Eq. \ref{eq:Mbh}); a $10^{10}\,\msun$ halo has a central black hole of $\approx 5.6\times10^{5}\,\msun$.  While these intermediate mass BHs are still theoretical, they can be built up by $z=6$ from $100\,\msun$ seeds at $z=20$ with a constant Eddington ratio of approximately 0.5.

%-------------------------------------------------------------------------------------------------------------
\subsection{Constraining Winds}\label{sec:UVLF:winds}

We want to consider parameter sets of our model that reasonably reproduce the M/L and its evolution with redshift determined by \citet{Munoz12} from fits to the UV LF at $z\gtrsim6$.  Since we expect the BH growth rate to be small compared to the galactic accretion rate, equation \ref{eq:Mdot1} becomes 
\begin{equation}\label{eq:SFR}
\dot{M}_{\rm disk} \approx (1+\eta_{\rm wind})\,\dot{M}_{\star}
\end{equation}
so that the SFR is set by the galactic accretion rate and the wind mass-loading parameter is given by equation \ref{eq:eta_wind}.  At the same time, a Salpeter IMF generates a luminosity per SFR of $8\times10^{27}\,{\rm ergs/s/Hz/(\msun/yr)}$ \citep[][but see the discussion in \citealt{ML11}]{Madau98}, however dust may diminish this intrinsic UV brightness.  To match the average observed luminosity of a $10^{10}\,\msun$ halo required by equation \ref{eq:Lave} ($\approx 1.6\times10^{27}\,{\rm ergs/s/Hz}$), we consider two possibilities: (1) that the amount of dust extinction is $\approx 0.18$ dex as determined by \citet{Bouwens07} from measurements of the UV continuum slope and (2) that $\eta_0=1$, where the difference between the resulting UV LF and that observed is a consequence of a flexible amount of dust extinction.  In both cases, we ignore the slight inconsistency between the presence of dust extinction and our dust-free pressure balance model from \S\ref{sec:model:acc_disk:model}.

The first case was considered by \citet{Munoz12}.  Since the amount of dust extinction at $z \approx 6$ is fixed, then for a given $\dot{M}_{\rm disk}$, $\eta_{0}$ is constrained to provide the required average luminosity in a $10^{10}\,\msun$ halo.  For the cold-flow accretion rate in equation \ref{eq:acc_cf}, the result is $\eta_{0}\approx 4$, where we have ignored the remaining slight mass dependence.  This is consistent with results from numerical simulations that require superwinds with $\eta_{0}=3$ (in our notation) to match observations at all redshifts in the {\it{WMAP3}} cosmology \citep{Dave06, OD08}.  If the average mass accretion rate were significantly larger, as may be the case in a scenario where $z \approx 6$ LBGs are identified with major mergers, the wind mass-loading factor would need to be correspondingly higher to produce the same average UV luminosity.

In the `flexible dust' case, we set $\eta_{0} = 1$ and calculate the dust extinction coefficient required to produce the correct observed luminosity.  Given this wind parameter at $z \approx 6$ in a cold-flow accretion model, a $10^{10}\,\msun$ halo must host galaxies that generate an average SFR of about $1\,{\rm \msun/yr}$.  This implies $\sim 0.68$ dex of dust absorption to produce the correct average observed UV luminosity inferred from the UV LF by \citet{Munoz12}.  Again, we note that the additional accretion rate in a merger scenario could be compensated for by a significantly increased amount of dust.  

The true physics involved in the formation and observation of high-redshift galaxies likely involves a combination of strong winds and some dust extinction.  While we adopt a fiducial value of $\eta_{0}=4$ in the rest of this {\it{Paper}}, we will also frequently show results for $\eta_{0}=1$ given the uncertainties in calculating the dust extinction from UV continuum slopes.  These values of $\eta_0$ correspond to critical wind velocities--below which winds remove more gas from the disk than does star formation--of $1200\,{\rm km/s}$ and $300\,{\rm km/s}$, respectively.  This implies mass-loading factors for $10^{10}\,\msun$ halos at $z=6$ of $\eta_{\rm wind} \approx 8$ and 2, respectively.\footnote{\label{foot:epsilon2}Although we have expressed the procedure in this section as fitting the parameter $\eta_0$ for fixed $\epsilon$, equation \ref{eq:SFR} shows that $\eta_{\rm wind}$ is really the more fundamental parameter.  Thus, as mentioned in footnote \ref{foot:epsilon1}, using a more realistic value of $\epsilon$ may result in different values $\eta_0$ because of the degeneracy between the two in equation \ref{eq:eta_wind}.  However, this will ultimately have no effect on the gas dynamics in our model.}

\begin{figure}
\begin{center}
\includegraphics[width=\columnwidth]{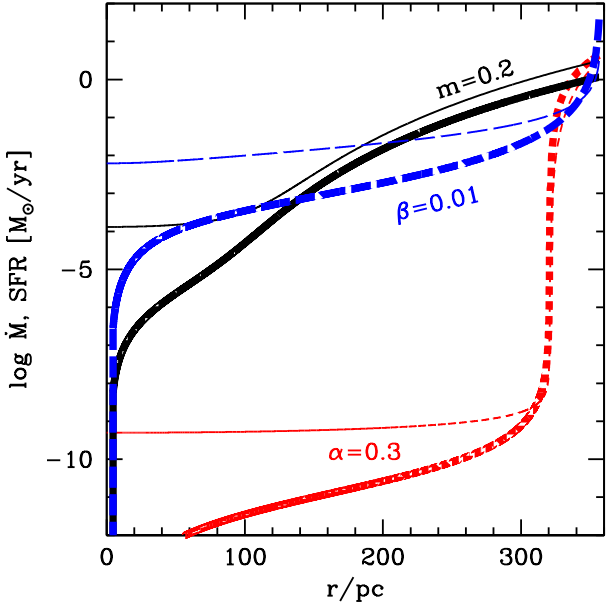}
\caption{\label{fig:mdot_sfr} 
The gas accretion rate (thin) and SFR ($\pi\,r^2\,\dot{\Sigma}_{\star}$; thick) as functions of radius in a galaxy hosted by a $10^{10}\,\msun$ halo at $z=6$ with $\eta_0=4$.  Short-dashed (red), solid (black), and long-dashed (blue) lines correspond to correspond to $\alpha$-disk, LSW, and shocked infall models with $\alpha=0.3$, $m=0.2$, and $\beta=0.01$, respectively.  
}
\end{center}
\end{figure}

%-------------------------------------------------------------------------------------------------------------
%             BH Accretion
%-------------------------------------------------------------------------------------------------------------
\section{The Black Hole Accretion Rate}\label{sec:BHacc}

As gas funnels toward the center of the disk, most is consumed by star formation or ejected by winds.  What remains is accreted onto the nuclear BH.  The rate at which gas is transported through the disk critically determines how much is available to feed the AGN.  The more rapid the transport, the more gas bypasses star formation and expulsion by winds.  We begin by numerically solving equation \ref{eq:Mdot1} and then investigate the qualitative behavior of approximate analytic solutions for each angular momentum transport model.  

Our numerical results show that, as expected, stronger winds and slower mass accretion velocities lead to steeper declines in the amount of gas transported to smaller radii.  If the decline is too steep (as in the $\alpha$-disk case), star formation uses up nearly all of the gas at the very outer edge of the disk resulting in a ring-like morphology in the UV.  Conversely, a more even star formation profile is achieved if the gas transport is rapid.  These results are shown in Figure \ref{fig:mdot_sfr}.  As the density of the disk increases towards its center, gas pressure begins to dominate stellar pressure.  No new stars form inside this radius, and the supply of gas inflow remains constant for $r \lesssim R_{\rm agn}$ so that $\dot{M}_{\rm BH} \approx \dot{M}(R_{\rm agn})$.   In Figure \ref{fig:mdot_sfr}, this occurs where the SFR turns down steeply at very small radii.  For $m=0.2$ in an LSW model with $\eta_0=4$, $R_{\rm agn}$ is on the order of a few parsecs.

\begin{figure*}
\begin{center}
\includegraphics[width=6.0in,trim=0 260 0 0,clip]{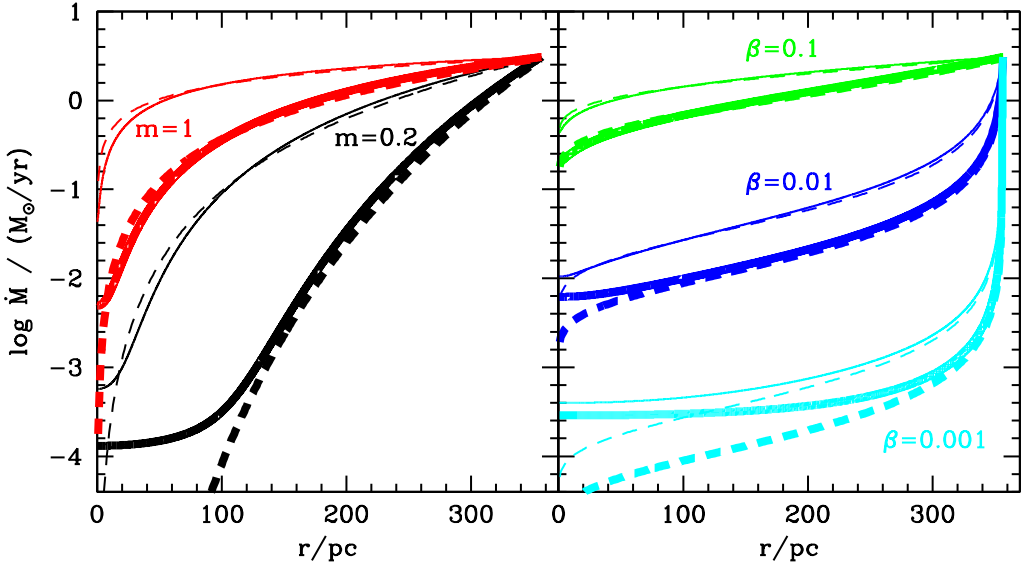}
\caption{\label{fig:mdot} 
The gas accretion rate as a function of radius in a galaxy hosted by a $10^{10}\,\msun$ halo at $z=6$.  Thin lines assume $\eta_{0}=1$, while thick lines assume $\eta_{0}=4$.  Solid curves show numerical results, while dashed curves show analytic solutions from Eq. \ref{eq:Mdot_sol_m} and \ref{eq:Mdot_sol_s} in the regime where pressure from stars dominates the hydrostatic equilibrium.  The left panel shows results for LSW infall with $m=0.2$ (black) and $m=1$ (red), while the right panel assumes shocked infall with blue, green, and cyan curves denoting $\beta=0.01$, $0.1$, and $0.001$, respectively.
}
\end{center}
\end{figure*}

As depicted in Figure \ref{fig:mdot_sfr}, $\alpha$-disk models, in which angular momentum transport is mediated by local viscosity, supply more than five orders-of-magnitude less gas to the AGN than LSW or shocked infall models for typical values of $\alpha$, $m$, and $\beta$.  We speculate that this dichotomy may represent two distinct modes of black hole growth.  Figure \ref{fig:mdot} shows that, for $M_{\rm halo}=10^{10}\,\msun$, an LSW model with $m=0.2$ and $\eta_0=1$ can build up the $5.6\times10^{5}\,\msun$ BH given by our assumed $M$-$\sigma$ relation (Eq. \ref{eq:Mbh}) by $z\sim6$ with a duty-cycle of about unity, where we define the duty cycle as the fraction of the age of the universe the AGN must be accreting at a constant rate to produce a BH of the given size.  The highest growth rate of the models we consider is produced by shocked infall with $\beta=0.1$ and $\eta_{0}=1$ and requires a low duty cycle of approximately 0.001.  The effect of variations in $m$ and $\beta$ for the LSW and shocked infall models, respectively, as well as changes in wind strength on the rate at which gas is transported through different parts of the disk can be seen in Figure \ref{fig:mdot}.  On the other hand, $\alpha$-disk models cannot come close to building up such a black hole without assistance from an additional phase of intense feeding.  Given the insignificant BH accretion rate produced by the local viscosity of an $\alpha$-disk, we consider only LSW and shocked infall models in subsequent sections of this {\it{Paper}}.  

The effects of varying the host halo mass or redshift on $\dot{M}_{\rm BH}$ are complex since they depend on the details of the transition to the AGN accretion disk.  However, numerical results in the LSW and shocked infall cases are shown in the top panels of Figure \ref{fig:agnM}.  While, the cold-flow galactic accretion rate at a given redshift increases roughly as $M_{\rm halo}$, the AGN accretion rate is sublinear with respect to $M_{\rm halo}$ in LSW and shocked infall models.

Finally, we note that our calculations ignore the effect of strong gravitational wave recoil during BH mergers.  In a scenario where high-redshift LBGs are the result of mergers, the newly merged BH may be kicked away from the dense center of the accretion disk and its growth rate correspondingly reduced \citep[e.g.,][]{Blecha08, TH09, Blecha11}.

%-------------------------------------------------------------------------------------------------------------
\subsection{Linear Spiral Wave}\label{sec:BHacc:mdot_LSW}

Exploring the behavior of the solution to equation \ref{eq:Mdot1} under some simplifying assumptions can give us qualitative insight into the influence of different angular momentum transport mechanisms on the BH accretion rate.  In the regime where $P_{\rm stars} \gg P_{\rm pgas}$, the vertical hydrostatic equilibrium (Eq. \ref{eq:pbal}) yields 
\begin{equation}\label{eq:S_SFR}
\dot{\Sigma}_{\star}=\frac{B\,\rho\,h^2\,\Omega^2}{2\,\epsilon\,c},
\end{equation}
with $\rho$ given by equation \ref{eq:rho}.  Additionally, combining equation \ref{eq:Mdot2} with an expression for the mass infall velocity gives the disk scale height as a function of radius that is required to funnel gas inward at a rate $\dot{M}$.  For an LSW model with an infall velocity given by equation \ref{eq:vin_m}, we find that $h/r \propto \dot{M}^{1/2}$.  Further substituting back into equation \ref{eq:Mdot1} yields 
\begin{equation}\label{eq:Mdot_m}
\dot{M}(r) \approx \dot{M}_{\rm disk}-A_{2}\,\int_r^{R_{\rm disk}}\frac{\dot{M}}{r'}\,dr',
\end{equation}
where
\begin{eqnarray}\label{eq:A2}
A_{2} & \equiv & \frac{B^{1/2}\,\sigma\,(1+\eta_{\rm wind})}{2\,\sqrt{2}\,\epsilon\,c\,m} \nonumber\\
& \approx & 8.3\,\left(\frac{\sigma}{50\,{\rm km/s}}\right)\,\left(\frac{1+\eta_{\rm wind}}{9}\right)\,\left(\frac{m}{0.2}\right)^{-1}.
\end{eqnarray}
Equation \ref{eq:Mdot_m} is an integral equation for $\dot{M}$ whose solution is
\begin{equation}\label{eq:Mdot_sol_m}
\dot{M}(r) \approx \dot{M}_{\rm disk}\,\left(\frac{r}{R_{\rm disk}}\right)^{A_{2}}.
\end{equation}
This solution was derived in Appendix D of TQM05, where their value of $A_{2}$ differs from ours because of our inclusion of winds and the disk self-gravity.  

Equation \ref{eq:Mdot_sol_m} reveals how the input parameters of our model influence the general behavior of the disk.  The infall rate is such that gas is available for star formation throughout the disk, as shown in Figure \ref{fig:mdot_sfr}.  The $\dot{M}$ drops off less steeply and the BH grows more rapidly for smaller values of $A_{2}$.  As we found numerically, this is achieved for smaller values of $\eta_{0}$ (i.e., less expulsion by winds) and larger values of $m$ (i.e., faster infall).  Moreover, including the self-gravity of the disk also results in slower BH growth since more gas must be turned into stars to maintain hydrostatic equilibrium, while the stronger supernovae increase the BH growth rate by allowing a lower SFR to produce the same hydrostatic pressure.  The dependences of the BH growth rate on halo mass and redshift are complicated since they enter through $\dot{M}_{\rm disk}$, $R_{\rm disk}$, and $R_{\rm agn}$, as well as the precise value of $A_{2}$.

We perform the above procedure in \S\ref{sec:BHacc:mdot_alpha} and \S\ref{sec:BHacc:mdot_shock} for $\alpha$-disk and shocked infall models of angular momentum transport.  

%-------------------------------------------------------------------------------------------------------------
\subsection{$\alpha$-disk}\label{sec:BHacc:mdot_alpha}

For an $\alpha$-disk model in the regime where pressure from stars dominates the hydrostatic equilibrium, equations \ref{eq:Mdot2} and \ref{eq:vin_a} combine to give $h/r \propto \dot{M}^{1/2}$, and equation \ref{eq:Mdot1} becomes 
\begin{equation}\label{eq:Mdot_a}
\dot{M}(r) \approx \dot{M}_{\rm disk}-A_{1}\,\int_r^{R_{\rm disk}}\frac{\dot{M}^{2/3}}{r'}\,dr',
\end{equation}
where
\begin{eqnarray}\label{eq:A1}
A_{1} & \equiv & \frac{B^{5/6}\,\sigma^2\,(1+\eta_{\rm wind})}{\sqrt{2}\,\epsilon\,c\,G^{1/3}\,Q^{1/3}\,\alpha}\nonumber\\
& \approx & 74\,{\rm \msun^{1/3}\,yr^{-1/3}}\,\left(\frac{\sigma}{50\,{\rm km/s}}\right)^2\,\left(\frac{1+\eta_{\rm wind}}{9}\right)\,\left(\frac{\alpha}{0.3}\right)^{-1},
\end{eqnarray}
and $\eta_{\rm wind}$ is given by equation \ref{eq:eta_wind}.  Differentiating both sides of equation \ref{eq:Mdot_a} with respect to $r$ and solving the resulting separable differential equation for $\dot{M}$, we find
\begin{equation}\label{eq:Mdot_sol_a}
\dot{M}(r) \approx \dot{M}_{\rm disk}\,\left[ 1 - \frac{A_{1}\,{\rm ln}\left(R_{\rm disk}/r\right)}{3\,\dot{M}_{\rm disk}^{1/3}} \right]^3.
\end{equation}
Because of the cubic power on the right-hand-side of equation \ref{eq:Mdot_sol_a}, the value of $\dot{M}$ drops quickly when $r<R_{\rm disk}$, as seen in Figure \ref{fig:mdot_sfr}.  The result is a ring-like galaxy morphology where all of the star formation takes place in the very outer parts of the disk in contrast to the smoother star formation distribution of the LSW case.  As $r \rightarrow 0$ in equation \ref{eq:Mdot_sol_a}, $\dot{M} \rightarrow -\infty$.  Of course, at some radius, the pressure from stars no longer dominates, and the assumptions under which we derived equation \ref{eq:Mdot_a} break down.  Indeed, $\dot{M} \geq 0$ at every radius in our model.  However, the unbounded nature of equation \ref{eq:Mdot_sol_a} highlights the rapid decline of $\dot{M}$ with decreasing radius.  Larger values of $A_{1}$ result in an even steeper decline.  Consequently, the expulsion of more gas via winds (i.e., larger $\eta_{0}$) results in slower BH growth, while a faster momentum transport rate (i.e., higher $\alpha$) allows gas to bypass star formation and winds and increases the BH accretion rate, as expected.

%-------------------------------------------------------------------------------------------------------------
\subsection{Shocked Infall}\label{sec:BHacc:mdot_shock}

Finally, in a shocked infall model, $h/r \propto \dot{M}$, and 
\begin{equation}\label{eq:Mdot_s}
\dot{M}(r) \approx \dot{M}_{\rm disk}-A_{3}\,\int_r^{R_{\rm disk}}\frac{\dot{M}^{2}}{r'}\,dr',
\end{equation}
where
\begin{eqnarray}\label{eq:A3}
A_{3} & \equiv & \frac{B^{1/2}\,G\,Q\,(1+\eta_{\rm wind})}{16\,\sqrt{2}\,\epsilon\,c\,\sigma^2\,\beta^2} \nonumber\\
& \approx & 70.\,{\rm \msun^{-1}\,yr}\,\left(\frac{\sigma}{50\,{\rm km/s}}\right)^{-2}\,\left(\frac{1+\eta_{\rm wind}}{9}\right)\,\left(\frac{\beta}{0.01}\right)^{-2},
\end{eqnarray}
and the solution to equation \ref{eq:Mdot_s} is
\begin{equation}\label{eq:Mdot_sol_s}
\dot{M}(r) \approx \frac{\dot{M}_{\rm disk}}{1+A_{3}\,\dot{M}_{\rm disk}\,{\rm ln}\left(R_{\rm disk}/r\right)}.
\end{equation}
Because of the large range of potential values of $\beta$ in shocked infall, the availability of gas to form stars inside the outer edge of the disk depends on the value of $A_{3}$.  When $A_{3} \gg 100$, the amount of available gas falls off steeply with decreasing radius (though, it's not as sudden a drop as in the $\alpha$-disk case) and a ring of star formation develops at the outer edge.  On the other hand, the more rapid infall rate of a disk with $A_{3} \ll 100$ leads to a smooth, potentially flat SFR profile and a higher BH growth rate.  As we have already seen, BH growth is accelerated by an increase in the infall rate (i.e., higher values of $\beta$) or a decrease in the expulsion rate from winds (i.e., smaller $\eta_{0}$).

%-------------------------------------------------------------------------------------------------------------
%             AGN
%-------------------------------------------------------------------------------------------------------------
\section{The AGN Luminosity}\label{sec:agn}

We now turn our attention to the luminosity generated by the BH accretion discussed in \S\ref{sec:BHacc}.  We consider the BH radiative efficiency and bolometric X-ray correction (\S\ref{sec:agn:xrays}) as well as obscuration along the line of sight (\S\ref{sec:agn:obscuration}).

\begin{figure}
\begin{center}
\includegraphics[width=\columnwidth]{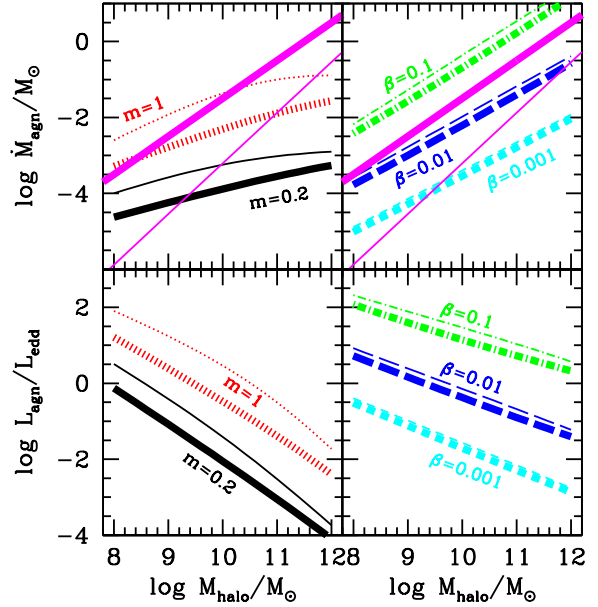}
\caption{\label{fig:agnM} 
The AGN accretion rate (top panels) and Eddington ratio (bottom panels) generated by the BHs in our model at $z=6$ as functions of host halo mass.  Thick and thin lines assume $\eta_0=4$ and 1, respectively.  Solid (black) and dotted (red) lines in the left-hand panels show results for LSW models with $m=0.2$ and 1, respectively, while long-dashed (blue), dot-dashed (green), and short-dashed (cyan) curves denote shocked infall models with $\beta=0.01$, 0.1, and 0.001, respectively.  For reference, the thick magenta line marks $M_{\rm halo} \propto \dot{M}_{\rm agn}$, while the thin magenta line shows the accretion rate required to build up a black hole commensurate with our assumed $M$-$\sigma$ relation by $z=6$ with a duty cycle of unity.
}
\end{center}
\end{figure}

\subsection{Intrinsic X-rays}\label{sec:agn:xrays}

The lost gravitational energy of the infalling gas in the inner accretion disk powers the central AGN so that the bolometric flux emitted at each radius is
\begin{equation}\label{eq:agn_flux}
F_{\rm grav}(r)=\frac{3\,\dot{M}\,\Omega^2}{8\,\pi}\,\left(1-\sqrt{\frac{R_{\rm in}}{r}}\right)
\end{equation}
If we take the radius of the inner edge of the disk to be that of the innermost stable circular orbit (ISCO) around a Schwartzchild black hole, $R_{\rm in}=6\,G\,M_{\rm BH}/c^2$, then the radiative efficiency of the AGN, $\epsilon_{\rm agn}=L_{\rm agn}/(\dot{M}_{\rm agn}\,c^2)$, is a constant $\epsilon_{\rm agn} \approx 8.33\%$ independent of any other model parameter.\footnote{This result ignores a general relativistic correction of order unity to the radiative efficiency.}  For a rotating central black hole, the radiative efficiency may be as high as tens of percent.

Using this efficiency factor, we calculate the bolometric AGN luminosity, $L_{\rm agn}$ and the Eddington ratios shown in the bottom panels of Figure \ref{fig:agnM}.  While models with very fast angular momentum transport seem to produce super-Eddington radiation, especially in small halos where the central BHs are smaller, this situation is probably unphysical.  These ratios could be achieved if the emission were collimated into jets, but it is more likely that the radiative efficiency would drop below that calculated above as outgoing photons are trapped by the infalling gas and ultimately dragged into the BH or that the radiation pressure would drive an AGN wind.

While the bolometric AGN luminosity is proportional to the AGN accretion rate given a constant radiative efficiency, X-ray luminosity is generated by Compton scattering processes that are difficult to predict from first principles.  The calculation becomes more tractable if we assume a typical AGN spectrum to determine the X-ray flux from the bolometric luminosity.  While a handful of quasars have been discovered at $z \gtrsim 6$ \citep[e.g.,][]{Fan06, Mortlock11}, these objects are quite rare.  Since substantial populations have not been built up to develop reliable template spectra at high redshift, we will instead deviate from our principle of not calibrating based on local observations and assume the bolometric correction for X-ray luminosity given by \citet{Marconi04}.  Where necessary, we adjust the given correction values of a particular band--for either a slightly different energy range or a band corresponding to the rest-frame at a higher redshift--by assuming an X-ray spectral slope of 1.9.

\subsection{Obscuration}\label{sec:agn:obscuration}

Because high-redshift galaxies are much denser than their low-redshift counterparts, the X-ray luminosities may be absorbed by high column densities of hydrogen.  Here we consider the amount of obscuration in terms of the fraction of impeded sight lines and the amount of gas available.  To give an approximate description of the central accretion disk properties, we assume that vertical hydrostatic equilibrium is maintained by radiation pressure in the single-scattering limit (see TWM05) generated by gravitational energy lost via accretion and no longer constrain $Q=1$.

If X-rays originate from a radius of order the inner edge of the disk, $R_{\rm in}$, then the fraction of obscured sight lines is approximately given by $h_{\rm in}/R_{\rm in}$, where $h_{\rm in}$ is the disk scale height at $R_{\rm in}$.  While a variety of model parameter sets can lead to geometrically thick disks in the TQM05 formalism \citep{Ballantyne08}, we find that all of our model disks are extremely thin in their centers.  However, even in a model with $h_{\rm in}/R_{\rm in} \ll 1$, sufficiently high X-ray luminosities from the central source can generate a obscuring wind from a dusty disk \citep{Chang07}.  Despite the fact that we have employed a dust-free framework for our calculations, our SFRs and BH accretion rates would be unchanged if we added a significant amount of dust to the inner BH accretion disk.  Given the contrast between the metals and dust observed in all $z \sim 6$ quasars \citep[e.g.,][]{Wang07, Wang11a, Wang11b} and the blue spectral slopes of the stellar continua \citep{Bouwens10a}, this two-zone dust model may indeed be appropriate, though of course, the galaxies hosting observed quasars are much more massive than the LBGs considered here.  Indeed, even with a dusty central disk, the intrinsic X-ray luminosities generated in our models are well below the level required to produce an inflated, unbound disk photosphere \citep{Chang07}.  Thus, we do not expect this mechanism to significantly increase the fraction of obscured sight lines.

Nevertheless, to give a sense of its potential effect, we consider a generous, toy-model estimate of X-ray obscuration by neutral hydrogen.  In this model, we assume that X-ray production proportional to the locally emitted gravitational energy is distributed throughout the plane of the BH accretion disk and must escape up through a local disk scale height, $h$.  At each radius, the X-rays are spectrally filtered by an energy-dependent factor of $e^{-S\,\sigma_{\rm H}/(2\,m_{\rm p})}$, where $\sigma_{\rm H}=10^{-17}\,{\rm cm^2}\,(E/13.6\,{\rm eV})^{-3}$ is the hydrogen photo-ionization cross-section.  While the bolometric correction of \citet{Marconi04} is calculated based on the {\emph{total}} bolometric emission and, therefore, does not apply to the emission at a particular radius, we set the radially-dependent bolometric correction to be the same in both the extincted and un-extincted cases and assume a fixed cross-section for each rest-frame band calculated from the central band energy, $\bar{E}$.  The resulting effective column density is
\begin{equation}\label{eq:coldens}
N(\bar{E})=\frac{1}{\sigma_{\rm H}(\bar{E})}\,{\rm ln}\left(\frac{\int_0^{R_{\rm agn}} F_{\rm grav}\,r\,dr}{\int_0^{R_{\rm agn}} e^{-\Sigma_{\rm g}\,\sigma_{\rm H}(\bar{E})/(2\,m_{\rm p})}\,F_{\rm grav}\,r\,dr}\right),
\end{equation}
where the intrinsic emission in a band with average energy $\bar{E}$ is suppressed by a factor of $e^{-N(\bar{E})\,\sigma_{\rm H}(\bar{E})}$.  We find that a significant amount of X-ray emission generated in the plane of the AGN disk is extincted by the dense gas.  Only a fraction $\sim 10^{-4}$--$0.1$ of the emitted X-rays in the 3.5--14${\,\rm keV}$ band escape to the observer.  The precise amount depends on the model parameters, and the exponential dependence on column density makes the range large.  In general, higher values of $m$ or $\beta$, while generating larger BH accretion rates and Eddington ratios (see Fig. \ref{fig:agnM}), also result in larger column densities.  We note that adopting $\eta_{0}=1$ affects the column density only minimally in comparison to changes in the angular momentum transport rate.  We note that the effective column density given in equation \ref{eq:coldens} is significantly higher than the average surface density of gas within the accretion disk, i.e., $\Sigma/(2\,m_{\rm p})$.  The difference between $\Sigma/(2\,m_{\rm p})$ and $N(\bar{E})$ is analogous to the difference between the average and effective optical depths in studies of the Ly$\alpha$ forest.  If the AGN accretion disk were uniform in density and brightness, these two quantities would be equal.  As it is, the increase in $F_{\rm grav}$ and $S$ toward the disk center also make the results independent of the precise value of $R_{\rm agn}$.

In the following section, we compare this highly obscured case to the case of unobscured emission appropriate if production of X-rays is near $R_{\rm in}$ and the disk is either geometrically thin or viewed face-on.

%-------------------------------------------------------------------------------------------------------------
%             Results
%-------------------------------------------------------------------------------------------------------------
\section{Results}\label{sec:results}

In the previous sections, we built up and calibrated a model that allows us to calculate galaxy disk properties, SFRs, and X-ray luminosities as functions of host halo mass and redshift for choices of the angular momentum transport mechanism and wind parameters.  In this section, we compare the resulting X-ray emission from $z \gtrsim 6$ LBGs with that produced by star formation (\S\ref{sec:results:hmxb}) and with {\it{Chandra}} observations (\S\ref{sec:results:xrays}).

%-------------------------------------------------------------------------------------------------------------
\subsection{Comparison with HMXBs}\label{sec:results:hmxb}

\begin{figure}
\begin{center}
\includegraphics[width=\columnwidth]{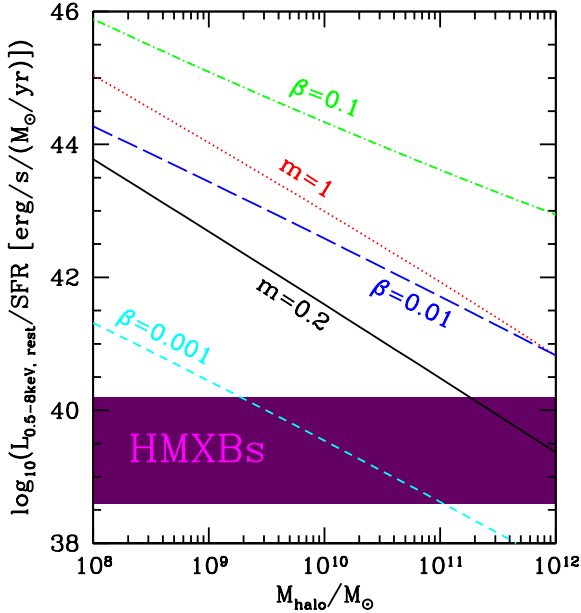}
\caption{\label{fig:Lx_sfr} 
The 2--$8\,{\rm keV}$ (rest-frame) X-ray luminosity to SFR ratio at $z=6$ as a function of host halo mass with $\eta_{0}=4$.  As in Fig. \ref{fig:agnM}, solid (black), dotted (red), long-dashed (blue), dot-dashed (green), and short-dashed (cyan) curves denote results for LSW models with $m=0.2$ and $m=1$ and shocked infall models with $\beta=0.01$, 0.1, and 0.001, respectively.  The shaded, magenta region marks the 2$\sigma$ range of reasonable values for $c_{\rm X}$ (representing emission from HMXBs) consistent with low-redshift observations of galaxies without AGN.  In general, most of our angular momentum transport models produce a significant AGN X-ray excess intrinsically over the expected emission from HMXBs.
}
\end{center}
\end{figure}

We first compare the X-ray luminosity from AGN in $z=6$ LBGs to that owing to HMXBs, the dominant mechanism for X-ray production in star-forming galaxies without AGN at low redshift \citep[e.g.,][]{Grimm03}.  Since HMXBs are short-lived, this mechanism is directly tied to the galactic SFR able to replenish them, a quantity straightforwardly calculated in our model.  The effect is usually parameterized by the ratio of X-ray luminosity to SFR: $c_{\rm X}=L_{\rm X}/{\rm SFR}$.  Measurements of the X-ray background and of individual galaxies at low redshift find a reasonable range for $c_{\rm X}$ such that the distribution of possible values is log-normal with a mean of $\left<{\rm log}\,c_{\rm X}\right>=39.4$ and a standard deviation of 0.4 \citep{Mineo12, Dijkstra12}.  If the AGN in LBGs at $z \gtrsim 6$ produce a significant amount of X-rays per SFR compared to HMXBs, it could affect the heating and ionization-state of the IGM and the 21 cm radiation signal during reionization \citep[e.g.,][]{Oh01, Furlanetto06}.  

Figure \ref{fig:Lx_sfr} shows the ratio of X-ray luminosity to SFR generated by the AGN in our models compared to the estimated range of $c_{\rm X}$ for HMXB consistent with low-redshift observations.  We have taken $L_{\rm X}$ in the 0.5--$8\,{\rm keV}$ (rest-frame) band, which we obtained by combining results using bolometric corrections for the 0.5--$2\,{\rm keV}$ and 2--$8\,{\rm keV}$ bands.  We find that most models produce intrinsic X-rays from AGN in excess of that from HMXBs except in the $\beta=0.001$ shocked infall model and the most massive halos in the $m=0.2$ LSW model.  These comparisons, of course, assume that the IMF and the resulting $c_{\rm X}$ from HMXBs do not vary significantly at high redshift from their low-redshift values.  If $z \gtrsim 6$ galaxies have lower metallicities, the contribution to the X-ray emission from star formation could be higher than expected \citep[e.g.,][]{Bookbinder80, Dray06, Linden10, Kaaret11}.  Additionally, the intrinsic X-ray emission from HMXBs in the star formation disk and from a central AGN may be obscured in very different ways and to different degrees.

%-------------------------------------------------------------------------------------------------------------
\subsection{X-rays from High-z LBGs}\label{sec:results:xrays}

\begin{figure*}
\begin{center}
\includegraphics[width=6.0in,trim=0 260 0 0,clip]{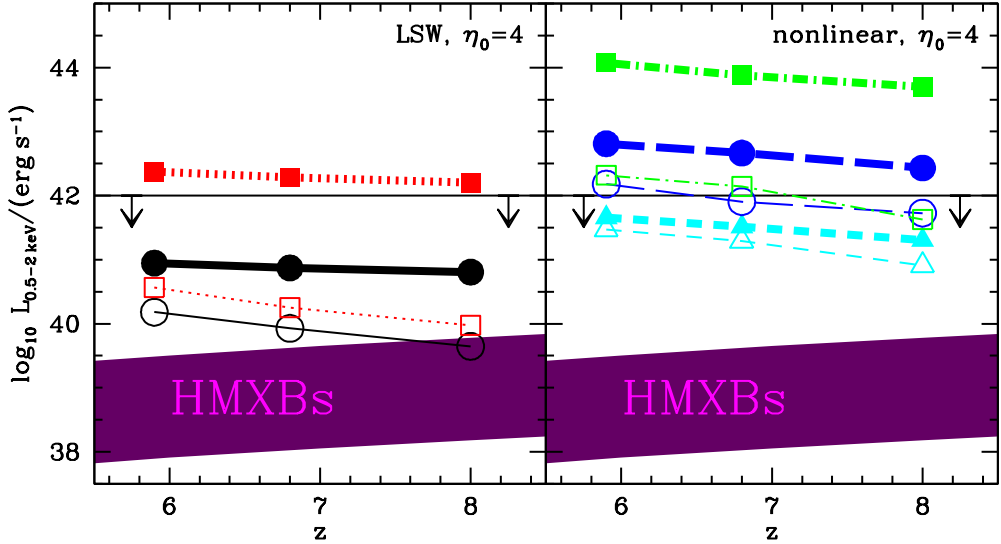}
\caption{\label{fig:Lx} 
Average X-ray luminosity ($0.5$--$2\,{\rm keV}$ observed frame; approximately $3.5$--$14\,{\rm keV}$ rest-frame) for the \citet{Bouwens07} and \citet{Bouwens11b} LBG samples at $z=5.9$, 6.8, and 8.0.  Thick lines connecting filled symbols denote unobscured values and are contrasted with maximally-obscured luminosities shown in thin lines connecting open symbols.  In the left panel, dotted (red, squares) and solid (black, circles) lines correspond to LSW models with $m=1$ and $m=0.2$, respectively.  In the right panel, dot-dashed (green, squares), long-dashed (blue, circles), and short-dashed (cyan, triangles) denote nonlinear shocked infall models with $\beta=0.1$, 0.1, and 0.001, respectively.  All of the calculations shown assume $\eta_0=4$ as since effect setting $\eta_0=1$ is much less than the difference between angular momentum transport mechanisms or between the maximally-obscured and unobscured cases.  The shaded, magenta region marks the X-ray luminosities (corresponding to a 2$\sigma$ range the distribution of $c_{\rm X}$) from a galaxy in a $10^{10}\,\msun$ halo due to the HMXBs resulting from star formation.  The horizontal, solid line shows the approximate \citet{Cowie12} upper limits.  
}
\end{center}
\end{figure*}

Analyzing recent {\it{Chandra}} observations \citep{Xue11}, \citet{Cowie12} placed upper limits on the X-ray signal from stacked populations of LBGs at $z\approx6$, 7, and 8 \citep{Bouwens07, Bouwens11b} in the soft, $0.5$--$2\,{\rm keV}$ (approximately $3.5$--$14\,{\rm keV}$ rest-frame) band.  They also argued against the claims by \citet{Treister11} of a strong detection at $z\approx6$.  

Using the probability distribution from equation \ref{eq:Mdist}, we randomly assign a halo mass to each galaxy in the LBG sample.  While the number of bright galaxies is not large enough to fully sample the halo mass distribution, their paucity also means that they contribute little to the average X-ray luminosity despite each one's being individually much brighter than any of the more common UV-faint LBGs.  We use these masses to calculate an X-ray luminosity (generated by our model AGN) for each LBG in the sample.  We show the average soft X-ray luminosities in Figure \ref{fig:Lx} compared with the \citet{Cowie12} upper limits at $z\approx6$--8.  Also plotted for comparison is an estimate of the HMXB contribution to the X-ray luminosity from galaxies hosted in $10^{10}\,\msun$ halos, where we have assumed an average SFR in these halos of $0.2\,{\rm \msun/yr}\,[(1+z)/7]^{2.5}$, normalized to the UV LF at $z=6$ and with the same mass scaling as the cold-flow accretion rate.  

We find that the average X-ray luminosity of $z \gtrsim 6$ LBGs is roughly consistent with the \citet{Cowie12} limits for all of our LSW models (left panel of Fig. \ref{fig:Lx}).  Additionally, all nonlinear-shock models also achieve consistency if X-ray production is distributed through the plane of the AGN accretion disk so that the disk itself obscures the emission.  By contrast, an unobscured, nonlinear accretion scenario only agrees with observations if $\beta \ll 0.01$.  The very rapid angular momentum transport rates of shock-induced accretion models produce, on average, about an order-of-magnitude more intrinsic X-ray emission than observed if $\beta=0.01$ and about two orders-of-magnitude more if $\beta=0.1$.  We have only plotted results for $\eta_0=4$ since setting $\eta_0=1$ produces smaller changes than varying $m$, $\beta$, or the amount of obscuration.  Indeed, in the obscured models, the precise values of $\eta_{0}$ and either $m$ or $\beta$, ultimately, have little effect on the resulting AGN luminosity.  This is because changing parameters to produce a higher BH accretion rate also tends to increase the absorbing column density.

%-------------------------------------------------------------------------------------------------------------
%       Conclusions
%-------------------------------------------------------------------------------------------------------------
\section{Conclusions}\label{sec:conclusions}

{\emph{We have extended the pressure-balanced ISM model of TQM05 to $z\gtrsim6$ using as few low-redshift or empirical prescriptions as possible.}}  In this model, gas accretes onto the outer edge of the galactic disk and is transported towards the center.  Along the way, star formation reduces the available gas in an amount necessary to maintain hydrostatic equilibrium and marginal Toomre-stability.  {\emph{Deviating from previous studies, we assumed a dust-free ISM, included the disk self-gravity and the expulsion of a significant fraction of the accreted gas by winds.}}  The galactic disk transitions smoothly into an accretion disk around a central BH providing the energy to power an AGN.  We calculate the physical and radiative properties of the disk as functions of radius by solving equation \ref{eq:pbal} under the constraints of equations \ref{eq:Mdot1} and \ref{eq:Mdot2} and assuming $Q=1$ in the outer portions of the disk.

We then used the UV LF analysis of \citet{ML11} and \citet{Munoz12} to calculate the the distribution of halo masses appropriate for $z\gtrsim6$ LBGs and calibrate the wind mass-loading factor.  {\emph{A cold-flow accretion rate can readily reproduce the observations with $\eta_{0}\approx4$}}, consistent with numerical simulations \citep{Dave06, OD08}.  However, where appropriate, we also explored the effect of assuming $\eta_{0}=1$ where consistency with the UV LF is restored by a somewhat larger amount of dust extinction than currently indicated by UV continuum slope measurements \citep{Bouwens07}.  For $10^{10}\,\msun$ halos at $z=6$, these two values of $\eta_{0}$ correspond to mass-loading factors of $\eta_{\rm wind} \approx 8$ and 2, respectively.  While we do not investigate in detail a scenario in which major mergers are responsible for $z \gtrsim 6$ LBGs, we expect that such a mechanism would generate an unaccountably high star formation or BH accretion rate when compared with observations.

A competition between the speed at which gas is transported toward the center of the disk and the rate at which it is transformed into stars or ejected by winds determines how much gas remains to be accreted onto the central BH.  Therefore, the distribution of star formation throughout the disk, the BH growth rate, and the resulting AGN luminosity critically depend on the angular momentum transport mechanism in the disk.  We compared the effect of three simple models: a local viscosity $\alpha$-disk, infall mediated by a linear spiral wave (LSW), and nonlinear, shocked infall from orbit-crossings.  We found that, while $\alpha$-disks cannot account for a growth rate high enough to form a black hole commensurate with our assumed $M$-$\sigma$ relation, {\emph{either an LSW model with $m=0.2$ and $\eta_0=1$ or a shocked infall model with $\beta \approx 0.001$ can build the correct size black hole by $z\sim6$ with a duty cycle of unity}} (see Fig. \ref{fig:agnM}).  We suggest that local, $\alpha$-disk models and globally-operating, gravitational torque models (such as LSW and shocked infall) may represent distinct growth modes for BHs in high-redshift LBGs.  Further investigation is required to determine whether the rapid angular momentum transport mediated by dense clumps and disk instabilities \citep{Bournaud11} can, in a full cosmological context, either stabilize or oscillate into a slower accretion mode dominated by local viscosity.

Given the dearth of $z\gtrsim6$ AGN, we calculated the X-ray bolometric correction by extrapolating from low-redshift observations.  However, the development of future samples will allow us to improve this procedure.  While the X-ray emission generated in $\alpha$-disk models is negligible due to minuscule BH accretion rates, {\emph{AGN in both the LSW and shocked infall models generate more X-ray luminosity than do HMXMs}}.  This is particularly true of shocked infall models for all reasonable input parameters (the exception being for $\beta=0.001$), while an unusually high value of $c_{\rm X}$ could allow HMXBs to compete with AGN in LSW models.  However, we also note that differences in the ways the AGN and the star-forming disk are obscured may complicate this comparison.  Future work, moreover, will show how this radiation feeds back into the IGM to modulate reionization.  

In considering the observable X-rays generated by our models, we considered two cases: unobscured emission--in which the emission is all produced at a radius of order the inner edge of the accretion disk and the disk is viewed face-on--and highly obscured emission--where X-ray production is distributed throughout the mid-plane of the accretion disk and radiation is spectrally-filtered through a disk scale height.  Interestingly, we find that our more extreme models are already ruled out by the current data; shocked infall models with $\beta \geq 0.01$ produce average intrinsic X-rays luminosities 1--2 orders-of-magnitude higher than observed.  This leads to the conclusion that {\emph{either angular momentum must be transported more slowly through the disk or the emission must be substantially obscured.}}
While the remaining models are all roughly consistent with the data, none of the shocked infall models (obscured or unobscured) is more than an order-of-magnitude below the limits.  The requirements of super-Eddington accretion and substantial obscuration may be circumstantial evidence arguing against non-linear, shocked infall in high-redshift LBGs.  However, {\emph{only moderately more sensitive observations are necessary to probe the more reasonable parameter-space.}}  While each individual X-ray observation need not get deeper, stacking a larger population of high-redshift LBGs might bear fruit.  Searches to expand the sample size of these objects are ongoing \citep{HIPPIES10, BoRG11, CANDELS11a, CANDELS11b}.  Focussing on lensed objects may also prove a way forward.  However, it is clear that exploring high-redshift galaxies outside of the traditional UV bands will provide interesting opportunities to verify and inform models of the internal physics and formation of these systems.  

%-------------------------------------------------------------------------------------------------------------
\section*{Acknowledgements}

We thank Eliot Quataert, Avi Loeb, and Molly Peebles for useful comments and discussions.  We also gratefully acknowledge a clear and critical review by the referee.  This research was partially supported by the David and Lucile Packard Foundation and by the Alfred P. Sloan Foundation.

%-------------------------------------------------------------------------------------------------------------

%-------------------------------------------------------------------------------------------------------------
\end{document}